\documentclass[pra,superscriptaddress,twocolumn,floatfix,showpacs]{revtex4-1}
\usepackage{graphicx,color}
\usepackage{amsmath,amssymb,bm}
\usepackage{braket}		

\newcommand{\mc}{\mathcal}

\newcommand{\fref}[1]{Fig.~\ref{#1}}
\newcommand{\eref}[1]{Eq.~\eqref{#1}}
\newcommand{\tref}[1]{Tab.~\ref{#1}}
\newcommand{\sref}[1]{Sec.~\ref{#1}}

\newcommand{\ahat}{\hat{a}^{\phantom{\dagger}}}
\newcommand{\ahatdag}{\hat{a}^{\dagger}}

\begin{document}
\title{Sudden expansion and domain-wall melting of strongly interacting bosons in  two-dimensional optical lattices and on multileg ladders}
\author{Johannes Hauschild}
\affiliation{Max Planck Institute for the Physics of Complex Systems, D-01187 Dresden, Germany}
\author{Frank Pollmann}
\affiliation{Max Planck Institute for the Physics of Complex Systems, D-01187 Dresden, Germany}
\author{Fabian Heidrich-Meisner}
\affiliation{Department of Physics and Arnold Sommerfeld Center for Theoretical Physics, Ludwig-Maximilians-Universit\"at M\"unchen, D-80333 M\"unchen, Germany}

\begin{abstract}
	We numerically investigate the expansion of clouds of hard-core bosons in the
	 two-dimensional square lattice using a matrix-product-state--based method.
         This nonequilibrium set-up is induced by quenching the trapping potential to zero
        and our work is specifically motivated by a recent experiment with interacting bosons in an optical lattice [Ronzheimer {\it et al.}, Phys. Rev. Lett. {\bf 110}, 205301 (2013)]. 
	As the anisotropy of the amplitudes $J_x$ and $J_y$ for hopping in different spatial directions is varied from the
        one- to the two-dimensional case,  
	we observe a  crossover from a fast ballistic expansion in the one-dimensional limit $J_x \gg J_y$ to
	much slower dynamics  in the isotropic two-dimensional limit $J_x = J_y$.  
	We further study the dynamics on multi-leg ladders and long cylinders.
	For these geometries we compare the expansion of a  cloud to the melting of a domain wall, which helps
        us to identify several different regimes of the expansion as a function of time.
	By studying the dependence of expansion velocities on both the anisotropy $J_y/J_x$ and the number of legs,
        we observe that the expansion on two-leg ladders, while similar to the two-dimensional case,  is slower than on wider ladders.
        We provide a qualitative explanation for this observation based on an analysis of the rung spectrum. 
\end{abstract}


\maketitle

\section{Introduction}

Ultracold quantum gases are famous for the possibility of realizing many-body Hamiltonians such as 
the Hubbard model, the tunability of interaction strength,
and, effectively, also dimensionality \cite{bloch08}. This provides access to genuine one-dimensional (1D) and 
two-dimensional (2D) physics as well as to the crossover physics between these limiting cases.
Moreover, time-dependent changes of various model parameters can be used to explore the nonequilibrium
dynamics of many-body systems (see \cite{langen15,gogolin15,eisert15} for recent reviews). Timely topics that are investigated in experiments include  the relaxation and thermalization dynamics in 
quantum quenches \cite{greiner02a,kinoshita06,hofferberth07,trotzky12,gring12,cheneau12,pertot14,will14,braun14,langen15a}, the realization of
metastable states \cite{weiss,vidmar15}, 
and nonequilibrium mass transport \cite{schneider12,reinhard13,ronzheimer13} and spin transport \cite{hild14}.
Due to the availability of powerful analytical and numerical methods such as bosonization \cite{giamarchi}, exact solutions for integrable systems \cite{essler-book}, or  the density matrix renormalization group method \cite{white92,schollwoeck05,schollwoeck11}, a direct comparison between theoretical
and experimental results is often possible in the case of 1D systems \cite{cheneau12,trotzky12,ronzheimer13,braun14}.

Strongly interacting many-body systems in two spatial dimensions, however,  pose 
many of the open problems in condensed matter theory and many-body physics, concerning both equilibrium 
and nonequilibrium properties. The reason is related to the lack of reliable numerical 
approaches.
Exact diagonalization, while supremely flexible, is inherently restricted to small system sizes \cite{rigol08}.
Nevertheless, smart constructions of truncated basis sets by selecting only states from subspaces that
are relevant for a given time-evolution problem have given access to a number of 2D nonequilibrium
problems (see, e.g., \cite{mierzejewski11,bonca12}). The truncation of equation of motions for operators provides an alternative approach \cite{uhrig09}, 
which has also been applied to quantum quench problems in the 2D Fermi-Hubbard model \cite{hamerla13}.
 Quantum Monte Carlo methods can be applied to systems in arbitrary dimensions including nonequilibrium problems (see, e.g., \cite{goth12,carleo12,carleo14}),
but suffer, for certain systems and parameter ranges, from the sign problem \cite{gull11}.  
Dynamical mean-field methods become accurate in higher dimensions, yet do  not necessarily yield quantitatively 
correct results in 2D \cite{eckstein10}. 

Regarding analytical approaches, we mention just a few examples, including solutions of the Boltzmann equation \cite{schneider12},
flow equations \cite{moeckel08}, expansions in terms of the inverse coordination number \cite{queisser14}, semiclassical approaches \cite{lux14,lux15}, or time-dependent mean-field approaches 
\cite{schutzhold06,schiro10,schiro11} such as the time-dependent Gutzwiller ansatz 
(see, e.g., \cite{jreissaty11,jreissaty13}). All these methods have provided valuable insights into aspects of the nonequilibrium dynamics in two (or three) dimensions, yet often involve
approximations.
Recently, the application of a nonequilibrium Green's function approach  to the dynamics in the sudden expansion in the 2D Fermi-Hubbard model has been explored \cite{schluenzen15}.

Although there have been very impressive  recent applications \cite{yan10,depenbrock12,stoudenmire12} of  the density-matrix renormalization group (DMRG) method \cite{white92} to 2D systems, 
the method, in general, faces a disadvantageous scaling with system size in 
2D \cite{stoudenmire12,schollwoeck05}. Tensor-network approaches  \cite{nishino01,verstraete04,jordan08} that were specifically designed to capture 2D many-body wave functions 
are an exciting development, with promising results for the $t-J$ model \cite{corboz11}.
A relatively little-explored area of research is the time evolution of 2D many-body systems in quantum quench problems using
DMRG-type algorithms \cite{zaletel14,Dorando2009,Haegeman2011,lubasch11,james15}.  

In this work, we present the application of a recent extension \cite{zaletel14} of 1D matrix-product state (MPS) algorithms \cite{vidal04,daley04,white04}
that is specifically tailored to deal with long-range interactions. Such long-range interactions arise 
by mapping even a short-range  Hamiltonian on a 2D lattice to a 1D chain for the application of DMRG.
 
Recent experiments have started to study the nonequilibrium dynamics of interacting quantum gases 
in 2D lattices or in the 1D-to-2D crossover \cite{schneider12,ronzheimer13,brown14}.
Motivated by Refs.~\cite{ronzheimer13,vidmar15}, we study the sudden expansion of hard-core bosons which is the release of
a trapped gas into a homogeneous optical lattice after quenching the trapping potential to zero.
The results of Ref.~\cite{ronzheimer13} show that strongly interacting bosons in 2D exhibit a much slower expansion
than their 1D counterpart. In the latter case, the integrability of hard core bosons leads to a strictly ballistic and (for the 
specific initial conditions of Ref.~\cite{ronzheimer13}) fast expansion that is indistinguishable from the one
of noninteracting fermions and bosons. In the 2D case, it is believed that diffusive dynamics sets in and virtually inhibits the
expansion in the high-density region, leading to a stable high-density core surrounded by ballistically expanding wings \cite{ronzheimer13},
similar to the behavior of interacting fermions in 2D \cite{schneider12}.
The characteristic feature of these diffusivelike expansions in contrast to the ballistic case is the emergence of
a spherically symmetric high-density core, while the ballistic expansion unveils the topology of the underlying reciprocal lattice.

\begin{figure}[tb]
\includegraphics{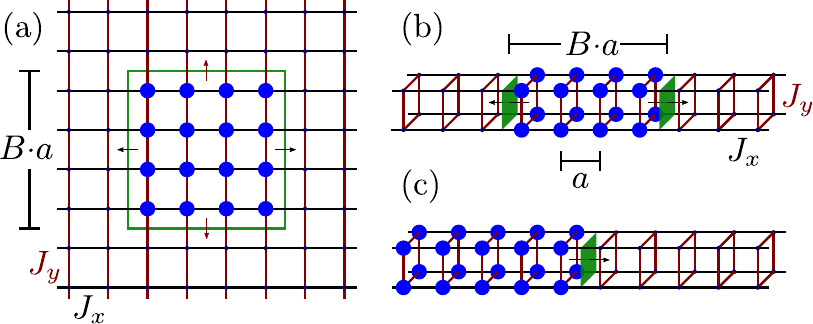}
\caption{(Color online) Illustration of initial states and geometries: (a) central block for the 2D expansion; (b)
        central block of size $ B\times L_y$; and (c) domain wall on a cylinder with $L_y=4$ legs.
}\label{fig:setup}
\end{figure}

In our work, we investigate this problem for both 2D clusters that can expand symmetrically in the $x$ and $y$ directions  [see Fig.~\ref{fig:setup}(a)] 
and wide cylinders and ladders [see Fig.~\ref{fig:setup}(b)]. We use the ratio of hopping matrix elements $J_x$ and $J_y$ along the 
$x$ and $y$ directions as a parameter to study the 1D-to-2D crossover. For the 2D expansion in the isotropic case $J_x=J_y$, 
we clearly observe the emergence of 
a spherically symmetric core, while for small values of $J_y<J_x$ and on the accessible time scales,
the expansion is essentially 1D-like. 
We further compute the expansion velocities derived from the time dependence of the radius as a function of $J_y/J_x$.

Since we are, in general, able to reach both longer times and larger particle numbers in the case of ladders than in 2D,
we present an extensive analysis of multileg ladders and cylinders [i.e., ladders with periodic boundary conditions in the 
(narrow) $y$ direction] with $L_y=2,3,4$ legs [see the sketch in Fig.~\ref{fig:setup}(b)]. From the analysis of the expansion in 1D systems \cite{vidmar15}, we expect that the
short-time dynamics is identical to the melting of so-called domain-wall states \cite{antal98,gobert05,lancaster10}, in which half of the system is empty
while the other  half contains one particle per site in the initial state [see the sketch in Fig.~\ref{fig:setup}(c)]. The domain-wall melting has been attracting considerable attention
as a nonequilibrium problem in 1D spin-$\frac{1}{2}$ systems (see, e.g., \cite{antal98,gobert05,lancaster10,caux11,santos11,sabetta13,halimeh14,alba14}).
Our results show that this similarity between the expansion of clusters and the domain-wall melting carries over to the transient dynamics on $L_y$-leg ladder systems, irrespective of boundary conditions.

A considerable portion of the discussion in both theoretical and experimental papers has focused on the question of whether
there are signatures of diffusive dynamics in the sudden expansion in 2D, in the dimensional crossover \cite{schneider12,ronzheimer13}, or on coupled chains \cite{vidmar13}.
The analysis of the expansion of fermions in the 2D square lattice starting from an initial state with two particles per site (i.e., a fermionic band insulator)
suggests that diffusive dynamics is responsible for the slow expansion in the high-density regions \cite{schneider12}.
This is expected to carry over to the bosonic case, yet there only two-leg ladders have been thoroughly studied.
In linear response, hard-core bosons on a two-leg ladder realize a textbook diffusive conductor at high temperatures \cite{steinigeweg14,karrasch15},
thus suggesting that diffusion may also play a role in the sudden expansion \cite{vidmar13}. Curiously, the expansion velocities measured numerically for hard-core
bosons on a two-leg ladder exhibit a dependence on $J_y/J_x$ that resembles the experimental observations for the true 2D case \cite{vidmar13,ronzheimer13}.
Here we are able to provide a more refined picture. Our analysis unveils that the sudden expansion becomes faster by going from two-leg to three- or four-leg
ladders. We trace this back to the existence of heavy excitations on the two-leg ladder that are defined on a rung of the ladder and are inherited from the $J_x \ll J_y $ limit, 
which cannot  propagate
in first-order tunneling processes in $J_x/J_y$. Conversely, the three- and four-leg ladders possess single-particle-like excitations, which we dub propagating modes, that have a sufficiently low
mass to become propagating.  This picture provides an intuitive 
understanding of the emergence of slow mass transport in the sudden expansion in the initial stages of the time evolution, complementary to the discussion of diffusive versus ballistic dynamics.
The reasoning is similar to the role that doublons play for slowing down mass transport in the 1D Bose-Hubbard model \cite{muth12,ronzheimer13,vidmar13,boschi14,sorg14},
which has also been emphasized in the case of the Fermi-Hubbard model \cite{hm09,kajala11}.
Our results raise the question as to whether the expansion in both directions in 2D and the one-directional expansion on wide ladders 
and cylinders will result in the same dependence of expansion velocities on $J_y/J_x$ for large $L_y$.
It appears that the ladders and cylinders, at least for small $L_y$,  preserve some degree of one-dimensionality.
A possible scenario is that the expansion velocities in the $x$ direction will depend nonmonotonically on $L_y$ for a fixed value of $J_y/J_x$ 
if ever they become identical to the behavior on the 2D systems.
As a caution, we stress that long expansion times may be necessary to fully probe the effect of a  2D expansion at small $J_y\ll J_x$ since the 
bare time scale for charge dynamics in the $y$ direction is set by $1/J_y$, as pointed out in \cite{pollet14}.

Apart from the nonequilibrium mass transport of strongly interacting bosons, there are also predictions for the
emergence of nonequilibrium condensates at finite quasimomenta in the sudden expansion in a 2D square lattice. These predictions are based on
exact diagonalization for narrow stripes \cite{hen10}, as well as on the time-dependent Gutzwiller method \cite{jreissaty11,jreissaty13}.
The dynamical condensation phenomenon has first been discussed for 1D systems (where it actually is a quasicondensation \cite{rigol04}), 
where it was firmly established  from  exact numerical results \cite{rigol04,rigol05} and analytical solutions \cite{lancaster10}
(see also \cite{micheli04,daley05,rodriguez06,vidmar13}) and has recently been observed in an
experiment \cite{vidmar15}.
In the sudden expansion of hard-core bosons in 1D, the dynamical quasicondensation is a transient, yet long-lived phenomenon \cite{rigol04,vidmar13} as ultimately
the quasimomentum distribution function of the physical particles approaches the one of the 
underlying noninteracting fermions via the dynamical fermionization mechanism \cite{rigol05a,minguzzi05}.

It is therefore an exciting question whether a true nonequilibrium condensate can be generated in 2D.
Our results cannot fully clarify this point, yet we do observe a bunching of particles at certain nonzero momenta
in the quasimomentum distribution after releasing the particles whenever propagating modes as discussed above are present.
For the melting of domain walls, the occupation of most of these modes, at which a nonequilibrium condensation is allowed by energy conservation and at which a bunching occurs,
 saturates  at long expansion times. The notable exception are certain modes on the   $L_y=4$ cylinder. This behavior, i.e., the saturation  is markedly different
from the 1D case of hard-core bosons in the domain-wall melting, 
where the occupation continuously increases.  The reason for this increase  is that the semi-infinite, initially filled half of the system will indefinitely feed the quasicondensates \cite{lancaster10,vidmar15}.
As such an increase is a necessary condition for condensation, 
we interpret the  saturation  of occupations as an indication that either breaking the integrability of strictly 1D hard-core bosons or the larger phase space for scattering
in 2D inhibits the dynamical condensation of expanding clouds. 
However, even in those cases on the ladder, in which we do not see a saturation, the increase is
slower than the true 1D case, suggesting that coupling chains, in general, disfavors condensation.
Yet a decisive analysis of this problem 
will require access to larger particle numbers and times in numerical simulations or future experiments.
Note that multileg ladder systems can be readily realized with optical lattices, using either superlattices \cite{foelling07} or the more recent approach of using
a synthetic lattice dimension \cite{celi13,stuhl15,mancini15}. 
Using a synthetic lattice dimension \cite{celi13}, it is in principle possible to obtain cylinders,
i.e., periodic boundary conditions along the (narrow) $y$-direction.

The plan of this paper is the following. In Sec.~\ref{sec:model}, we introduce the model and definitions. Section~\ref{sec:radii}
provides a discussion and definitions for various measures of expansion velocities employed throughout our work, while Sec.~\ref{sec:num}
provides details on our numerical method.
We present our results for the 2D case in Sec.~\ref{sec:2Dexpansion}, while the results for multileg ladders and cylinders are contained
in Sec.~\ref{sec:legs}. We conclude with a summary presented in Sec.~\ref{sec:sum}, while details on the extraction of velocities and on the diagonalization of rung Hilbert spaces are  contained in two appendixes.

\section{Model and initial conditions}
\label{sec:model}

We consider hard-core bosons on a square lattice and on multileg ladders. 
The Hamiltonian reads
\begin{multline}
	H = - \sum_{i_x,i_y}\lbrack J_x  (\ahatdag_{i_x,i_y} \ahat_{i_x+1,i_y} +h.c.) \,   \\ 
	+ J_y (\ahatdag_{i_x,i_y} \ahat_{i_x,i_y+1} +h.c.) \rbrack\, .
	 \label{eq:ham}
\end{multline}
Here $\ahatdag_{i_x,i_y}$ denotes the creation operator on site $\mathbf{ i}=(i_x, i_y)$
and $J_x$($J_y$) are the hopping matrix elements in the $x$($y$) direction.
We choose the hopping matrix element $J_x$ in the $x$ direction and the lattice constant $a$ as units and set $\hbar$ to unity;
the ratio $J_y/J_x$ is dimensionless.
Note that the Hamiltonian is equivalent to the spin-$\frac{1}{2}$ $XX$ model. In 1D ($J_y=0$), the Jordan-Wigner
transformation maps the bosons to free fermions \cite{cazalilla11}. 
$L_x$ and $L_y$ denote the number of sites in the $x$ and $y$ direction, respectively.

We  consider different geometries, namely 
(i) a small square-shaped cluster of $L_x = L_y = 12$ sites with open boundary conditions in both directions,
(ii) ladders with $L_x = 60$, $L_y \in \set{2,3,4}$ with open boundary conditions (OBCs) in both the $x$- and $y$-direction, 
and (iii) cylinders with $L_x = 60$, $L_y \in \set{2,3,4}$ with periodic boundary conditions (PBCs) in
the $y$ direction and OBCs in the $x$ direction.  
For two-leg ladders, the only difference between the Hamiltonian with OPC and PBC along the $y$ direction is thus a factor of two in the tunneling matrix element $J_y$.
In praxis, we obtain the behavior with PBCs by just taking the OBCs data with $J_{y} \rightarrow J_y/2$.

For all simulations, we start the expansion from a product state, 
\begin{equation} | \psi_0 \rangle = \prod_{{\bf i} \in \mc{B}} \ahatdag_{i_x,i_y} \ket{\rm vac},
\end{equation} 
in real space.
To model the fully 2D expansion, we choose $\mc{B}$ to be a square-shaped block of
$B\times B$ sites centered in the cluster; see \fref{fig:setup}(a).
On cylinders and ladders, we study two different types of  $\mc{B}$: 
(i) a block of $B \times L_y$ bosons, centered in the $x$ direction and filling all the sites in
the $y$ direction as shown in \fref{fig:setup}(b),
and (ii) a domain wall, where the left half of the lattice is occupied by a block of $L_x/2
\times L_y$ bosons while the right half is empty; see \fref{fig:setup}(c).

\section{Definitions of expansion velocities}
\label{sec:radii}

There are several possible ways of defining the spatial extension of an expanding cloud and thus also several
different velocities. 

\subsection{Position of the fastest wave front} 
One can define the cloud size from its maximum extension, i.e., from the
position of the (fastest) wave front. The velocity derived from this approach will typically simply
be the fastest possible group velocity (provided the corresponding quasimomentum is occupied in the initial state).
Thus, this velocity will not contain information about the slower-moving particles and any emergent slow and possibly diffusive dynamics in
the core region. We do not study the wave front in this work.

\subsection{Radial velocity} 

Theoretically, it is natural to define the radius $R$ as the square root of the second moment of the
particle distribution $\langle n_i(t)\rangle $. Suppose we are interested in the expansion in $x$ direction: We average the density profile over the
$y$ direction to calculate the radius
\begin{equation}
	R_x^2(t) = \frac{1}{N} \sum_{i_x, i_y} n_{i_x,i_y}(t) (i_x a - i_x^0 a)^2, 
	\label{eq:R}
\end{equation}
where $i_x^0 a$ is the center of mass in the $x$ direction and $N$ is the total number of bosons.
An analogous expression is used to define $R_y^2$.
To get rid of an initial constant part, we use $\tilde{R}_\mu^2(t) =  R_\mu^2(t) - R_\mu^2(t=0) $ to
define the radial velocity 
\begin{equation}
	v_{r,\mu} = \frac{\partial \tilde{R}_{\mu}(t)}{\partial t}\, 
	\label{eq:vr}
\end{equation}
with $\mu=x,y$.
The corresponding velocity has contributions from
all occupied quasimomenta. It will ultimately be dominated by the fastest expanding particles, and for the
sudden expansion, $R_\mu$ will be linear in time in the limit in which the gas has become dilute and effectively noninteracting.

The radial expansion velocity of 1D systems was studied for the Fermi-Hubbard model \cite{langer12}, the Bose-Hubbard model \cite{ronzheimer13,vidmar13}, and the Lieb-Liniger model \cite{jukic09}.
For Bethe-integrable 1D systems, it can be related to distributions of rapidities \cite{mei15}.
For a recent study of the radial velocity in the 2D Fermi-Hubbard model, see \cite{schluenzen15}.
 
\subsection{Core expansion velocity} 
In the related experiments with ultracold atoms \cite{schneider12,ronzheimer13}, the focus was on the
core expansion velocity that is derived from the time evolution of the half width at half maximum $r_c(t)$.
The reason is that  in these experiments, 
 an  average over many 1D or 2D systems is measured. Moreover, the core expansion velocity is primarily sensitive
to the dynamics in the high-density core (but  insensitive to the ballistic tails) and thus yields slightly different information.
In case of multiple local maxima, the two outermost points are taken.
Since in our simulations we have smaller particle numbers compared to the   experiments \cite{schneider12,ronzheimer13},
we use linear
splines to interpolate the density profile between the lattice sites in order to get values for
$r_c(t)$ to a better  accuracy than just a single lattice constant.
The core expansion velocity is defined as the time derivative
\begin{equation}
 v_c = \frac{\partial r_c(t)}{\partial t} \,.
\end{equation}
The full time dependence of $r_c$ and the extraction of $v_c$ is discussed in Appendix \ref{sec:extractvcvr}.

\section{Numerical Method}
\label{sec:num}
Although the Hamiltonian \eref{eq:ham} itself is short ranged, long-range interactions arise by
mapping the 2D lattice to a 1D DMRG chain. The presence of such long-range interactions 
renders most of the existing DMRG-based algorithms for the time evolution \cite{vidal04,daley04,white04,schollwoeck11}  
inefficient because a direct Trotter decomposition of the exponential is not possible.
In our work, we use a recently developed extension of an MPS-based time-dependent DMRG algorithm 
that is particularly suited for such systems \cite{zaletel14}.
The method is based on a local version of a Runge-Kutta step which can be efficiently represented
by a matrix-product operator (MPO) \cite{Verstraete2004}.
The actual time evolution can then be performed using standard algorithms that apply an MPO to a given MPS \cite{schollwoeck11}.
An advantage of the method is that it can be easily implemented into an existing MPS based DMRG code and has a constant error per site.    
 
For our simulations, we choose the DMRG chain to wind along the $y$ direction in order to keep the range of the interactions as small as
possible (namely $L_y$).
Sources of errors are the discretization in time and the discarded weight per truncation of the MPSs after each time step. 
The time steps are  chosen small enough to make the error resulting from the second-order expansion negligible.
We furthermore choose the truncation error at each step to be smaller than $10^{-10}$, which is sufficient to obtain all measured observables accurately.
The  growth of the entanglement entropy following the quench requires increasing the bond dimension $\chi$ with time.
Conversely, since we  restrict the number of states to $\chi \lesssim 2000$,  we are naturally limited to a finite maximum time $t_m$ at which the truncation error becomes significant.
Note that the bond dimension $\chi$ required for the simulations grows exponentially with time.
Increasing the particle numbers and $L_y$ leads to a faster growth of the entanglement entropy and thus to a shorter maximal time $t_m$.
However, we stress that we clearly reach longer times and larger systems than is accessible with exact diagonalization (i.e., pure state propagation using, e.g., Krylov subspace methods).

\section{Two-dimensional expansion}
\label{sec:2Dexpansion}

\subsection{Density profiles}
\begin{figure}[tb]
\includegraphics[width=\columnwidth]{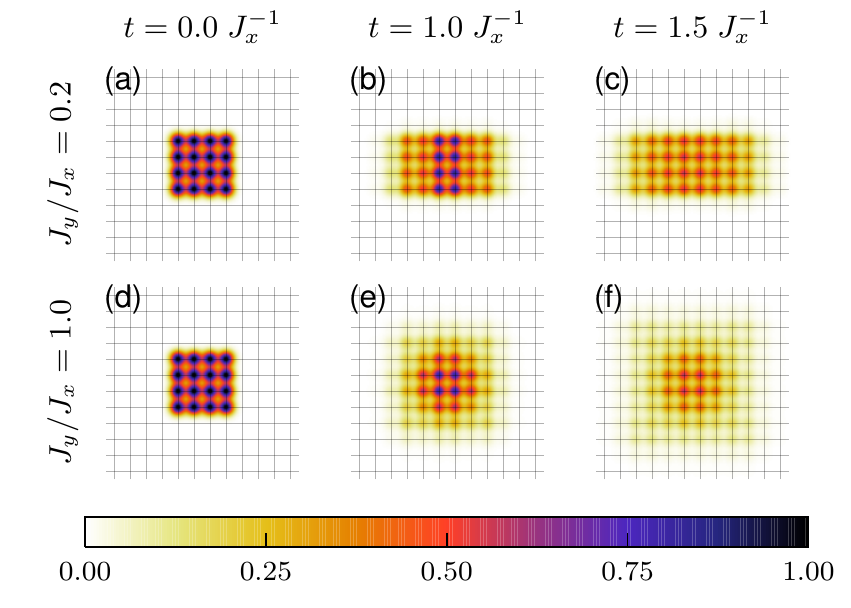}
\caption{(Color online) Density profiles for the 2D expansion from a $4\times4 $ cluster with 
	(a)--(c)	$J_y/J_x = 0.2$  and (d)--(f) $J_y/J_x = 1.0$ at times $t J_x =0.0 ,1.0,1.5 $.
}\label{fig:2Dprofile}
\end{figure}

We first characterize the expansion by analyzing  the time- and position-resolved  density profile
$n_{i_x,i_y} (t) = \braket{\hat{n}_{i_x,i_y}(t)}$, where $\hat{n}_{i_x,i_y}= \ahatdag_{i_x,i_y}
\ahat_{i_x,i_y}$ is the number operator. We present exemplary density profiles for three different times and 
and two anisotropies $J_y/J_x \in \set{0.2, 1}$ in \fref{fig:2Dprofile}.
For small $ J_y/J_x =0.2 $ [Figs.~\ref{fig:2Dprofile}(a)--(c)], there is  a fast expansion in the $x$ direction and nearly no
expansion in the $y$ direction. This is expected since the bare timescale for the expansion in the $y$ direction set by $1/J_y$ is
here much larger than the one in the $x$ direction \cite{pollet14}.
On the other hand, for $J_y= J_x$, we find four ``beams'' of faster expanding particles going out
along the diagonals.
These beams are even more pronounced for initial states with smaller clusters of $2\times2$ and
$3\times3$ bosons (not shown here).

The most important qualitative difference between the density profiles at $J_y/J_x=0.2$ and  $J_y/J_x=1$ is the {\it shape}.
In the former case, the profiles retain a rectangular form, reflecting the underlying reciprocal lattice and the different
bare tunneling times in the $x$ versus the $y$ direction. For the isotropic case, the initial square shape of the cluster changes into
a spherically symmetrical form in the high-density region. This observation is consistent with the experimental results of
\cite{ronzheimer13}.

\subsection{Radial velocity}
\begin{figure}[tb]
\includegraphics[width=\columnwidth]{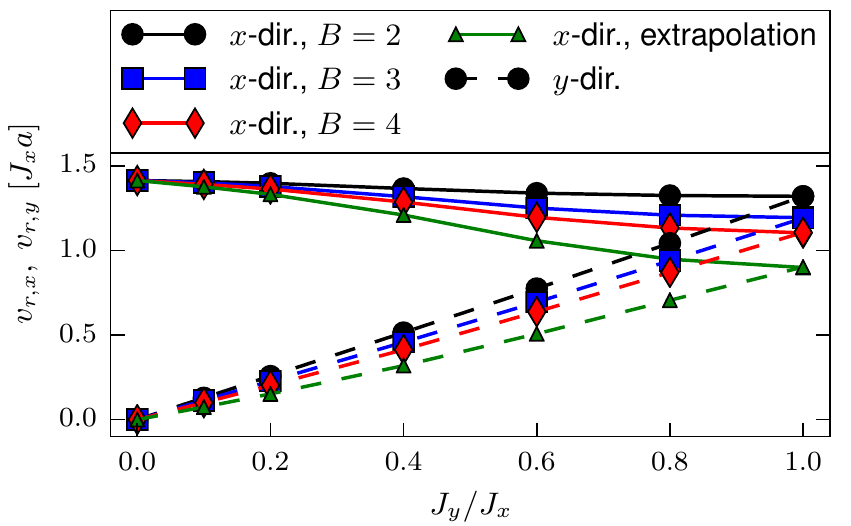}
\caption{(Color online) Radial velocity $v_{r,x/y}$ in the $x$ direction (top three solid lines) and the $y$ direction
	(dashed lines) for the 2D expansion from $B\times B $ clusters.
	The small green triangles show the result of an extrapolation to $B=\infty$ using \eref{eq:vrScaling}.
	}
\label{fig:2Dvr}
\end{figure}
In order to compare the expansion for different values of $J_y/J_x$ more quantitatively, 
we extract certain integrated quantities from the profiles, which contain relevant information.
One such quantity is the radial velocity $v_{r,x/y} $ derived from the reduced radius $\tilde R_{x/y}$ [see Eq.~\eqref{eq:R}].
Details on how we extract $v_r$ from the time-dependent reduced radius $\tilde{R}(t)$ can be found in  Appendix
\ref{sec:extractvcvr}.

The radial velocities $v_{r,x}$ and $v_{r,y}$ for the 2D expansion are shown in \fref{fig:2Dvr}.
Unfortunately, our simulations for the 2D lattice are restricted to both very short
times and small numbers of bosons with block sizes $B \in \set{2,3,4}$. For instance, for $4\times4$
bosons we reach only times $t_m \approx 1.5 \, J_x^{-1}$.
The short times prevent us from a reliable extraction of the core expansion velocity, which would allow
for a direct comparison to the experiment \cite{schneider12,ronzheimer13}.
The experimental results \cite{ronzheimer13} suggest that, for increasing $J_y$, the core expansion
velocity in the $x$ direction decreases 
dramatically (see \fref{fig:cylladvcvr}), which has been attributed to the breaking of integrability of 1D hard-core bosons \cite{ronzheimer13,vidmar13}.

Our results for the radial velocity $v_r$ show  that for the smallest block size $B=2$, tuning $J_y/J_x$ from 0 to 1 changes the velocity  $v_{r,x}$
only gradually
while the velocity in the $y$ direction scales almost linearly with
$J_y$.
A previous study of the expansion of two-leg ladders also indicated that the core expansion velocity exhibits a much stronger dependence on $J_y/J_x$ 
than the radial expansion velocity \cite{vidmar13}.
We suspect that this weak dependence may additionally result  from the  small number of bosons considered in our simulations:
Increasing $J_y$ allows a hopping in the $y$ direction, which reduces the density and thus the effective interaction.
In other words, tuning $J_y/J_x$ from 0 to 1 increases the effective surface of the initial
block to include the upper and lower boundaries. From the surface,  there is always a fraction of the bosons that  escape 
 and which effectively do not experience the hard-core interaction.
This effect becomes  more relevant for smaller boson numbers, where the bosons are almost
immediately dilute, feel no effective interaction, and, thus, expand (nearly) ballistically in both
directions.
For larger block sizes $B = 3,4$, the ratio of surface to bulk is smaller and, therefore,
interaction effects become more relevant. 
Indeed, we find for $B=3, 4$ that tuning $J_y/J_x$ from 0 to 1 leads to a significant reduction
of $v_{r,x}$, most pronounced for $B=4$.

Even though we have access to only
three values of $B$, it is noteworthy that for all values of $J_y/J_x$, $v_{r,x}(v_{r,y})$ decreases (increases) monotonically with $B$ 
and thus with total particle number. 
This tendency is compatible with the behavior of the experiments \cite{ronzheimer13} performed with
much larger boson numbers, which motivates us to perform an  extrapolation to $B=\infty$ despite the small number of bosons.
We assume that the finite-size dependence is dominated by the surface effects of the initial boundary,
which scales with $B$. Therefore, we extract the velocity for $B=\infty$ from a fit to the form
\begin{equation}
	v_{r,x/y}(B) = v_{r,x/y}(B=\infty) + \frac{\mathrm{const}}{B} \label{eq:vrScaling}
\end{equation}
at fixed $J_y/J_x$. 
 The  resulting values, which are indicated by the
small green symbols in \fref{fig:2Dvr}, should only be considered as rough estimates.

\subsection{Momentum distribution function}
\begin{figure}[tb]
	\includegraphics[width=\columnwidth]{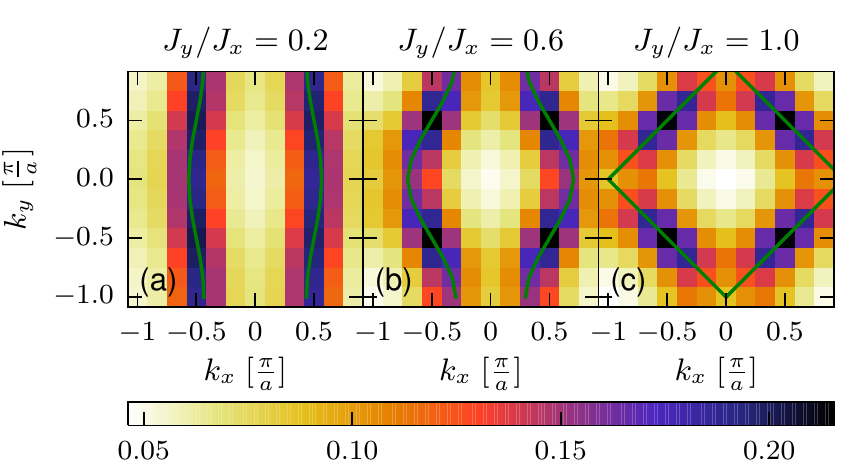}
	\caption{(Color online) Momentum distribution function $n_{k_x,k_y}$ (dimensionless) for the 2D expanding cloud of $4 \times 4 $ bosons at time $t = 1.5 \,J_x^{-1}$.
		The solid green lines show the solutions to \eref{eq:EnergyCondensates}.
	} 
	\label{fig:2Dmomentum}
\end{figure}
Figure \ref{fig:2Dmomentum} shows the momentum distribution function 
\begin{multline}
	n_{k_x,k_y} = \frac{1}{L_x L_y} \sum_{i_x, i_y, j_x, j_y} 
	e^{-\mathrm{i} ( k_x (i_x a-j_x a) + k_y(i_y a-j_y a))} 
 	\\
	\times \braket{\ahatdag_{i_x,i_y} \ahat_{j_x,j_y}}
\end{multline}
for the 2D expansion.
For a purely 1D expansion ($J_y=0$),  dynamical  quasicondensation occurs at $k_x = \pm
\frac{\pi}{2 a}$ \cite{rigol04,rigol05,vidmar15}.
As discussed in Refs.~\cite{hen10,jreissaty11}, energy
conservation restricts the (quasi)condensation to momenta at which the single-particle dispersion relation
$\epsilon(k_x,k_y)$ vanishes since the initial state has zero energy, resulting in the emission of bosons with, on average,  zero energy per particle. For a 2D system, this leads to 
\begin{align}
	\epsilon(k_x, k_y) &= -2 J_x \cos(k_x a) - 2 J_y \cos(k_y a) =0 . \label{eq:EnergyCondensates}
\end{align}
The solutions of this equation are   indicated by the solid green lines in \fref{fig:2Dmomentum}. We indeed observe an accumulation of particles
at momenta compatible with \eref{eq:EnergyCondensates}.
For $J_y/J_x = 0.2$ [\fref{fig:2Dmomentum}(a)], there is  almost the same weight at any 
momentum $k_y$ compatible with Eq.~\eqref{eq:EnergyCondensates}. We suspect that this  is a relict
of the short time $t = 1.5 \,J_x^{-1} = 0.3 \,J_y^{-1}$ reached in the simulations: Up to this time there was almost no expansion in the $y$ direction; thus, we
have roughly $\braket{\ahatdag_{i_x,i_y} \ahat_{j_x,j_y}} \approx \delta_{i_y, j_y}$ such that $n_{k_x,k_y} $
is initially independent of $k_y$.
Nevertheless, closer inspection shows slightly more weight at compatible momenta with $k_y=\pm
\frac{\pi}{2a}$ than at
those with $k_y = 0$ even for small $J_y$ [see \fref{fig:2Dmomentum}(a)]. 
This becomes much more pronounced for $J_y = J_x$ [see \fref{fig:2Dmomentum}(c)]. In this case, the strongest peaks are at
$(k_x,k_y) = (\pm \frac{\pi}{2a}, \pm \frac{\pi}{2a}), (\pm \frac{\pi}{2a}, \mp \frac{\pi}{2a})$. These four points correspond to the maximum group velocities
$v(k_x,k_y) = (2 J_x a \sin(k_x a), 2 J_ya \sin(k_y a) )$ and, in real space, manifest themselves via  the four ``beams'' in the density profile shown in \fref{fig:2Dprofile}(f).

Our results do not serve to clarify whether there actually is a dynamical condensation at finite momenta in 
2D or not since our initial clusters have too few particles in the bulk compared to their surface.
The fast ballistic propagation of the particles melting away from the surface will only be suppressed
once the majority of particles is in the bulk initially.
If we attribute the outermost particles to the surface, this would require us to be able to simulate at least $7 \times 7$ clusters. 
We believe that the accumulation at finite momenta seen
in the quasimomentum distribution function is due to these fast particles melting away from the boundary during the first tunneling time. 
Moreover, we would need to be 
able to study the particle-number dependence of the height of the maxima in the quasimomentum distribution function or the decay of single-particle
correlations over sufficiently long distances \cite{rigol04}.

\section{Cylinders and ladders}
\label{sec:legs}

In contrast to the 2D lattice, the ratio of surface to bulk is much lower for cylinders and
ladders, as we initialize the system uniformly in the $y$ direction.
Moreover, if we tune $J_y$ from 0 to 1, the additional hopping in the $y$ direction does not lower the
density (and with it the effective interaction), as it is the case for the fully 2D
expansion.
We thus expect a weaker dependence of the results on the number of bosons.
Additionally, we can reach larger times than for the fully 2D expansion since the
range of hopping terms after mapping to the DMRG chain is smaller.
While we can reach times up to $t_m \approx 6 \,J_x^{-1}$ for $L_y = 2$, we are restricted to times up to
$t_m \approx 4 \,J_x^{-1}$ for $L_y =3$ and $t_m \approx 3 \,J_x^{-1}$ for $L_y = 4$.

\begin{figure}[tb]
\includegraphics[width=\columnwidth]{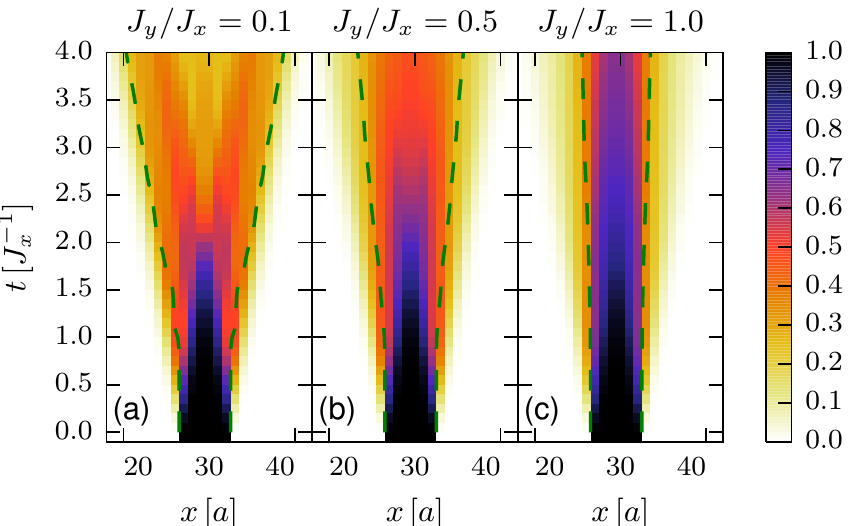}
\caption{(Color online) Integrated density profiles $\frac{1}{L_y}\sum_{i_y} n_{i_x,i_y}(t)$ (dimensionless) for the expansion
from a $6\times3 $ cluster on a cylinder with $L_y=3$. The green dashed lines show the location of the half
maximum on the left and right.
}\label{fig:cylprofilesr}
\end{figure}

\subsection{Density profile}
\label{sec:legs_density}
\begin{figure}[tb]
\includegraphics{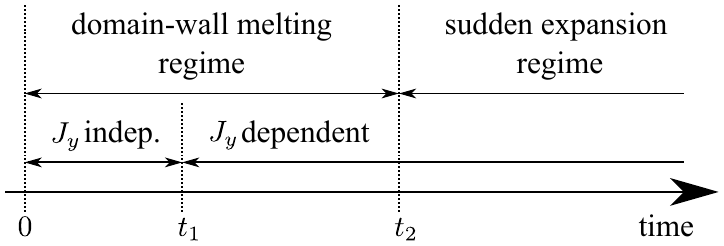}
\caption{Illustration of the time regimes for the expansion of blocks (see the text in Sec.~\ref{sec:legs_density} for details).}
\label{fig:timeregimes}
\end{figure}

Figure \ref{fig:cylprofilesr} shows some typical results for the  column density for the expansion
of a block on a cylinder with $L_y=3$.
We identify three different time regimes for the expansion of blocks, schematically depicted in \fref{fig:timeregimes}.
First, the evolution during the first tunneling time $t_1 \propto 1/J_x$ is independent of $J_y$:
Since we initialize our system uniformly in $y$ direction, in the initial longitudinal hopping, there cannot be any dependence on $J_y$
and a finite amount of time is required before 
correlations in the $y$ direction can build up.

Then, in a transient regime $0<t_2$ (where $t_2>t_1$), the melting of the block from either side is equivalent to the  domain-wall melting 
\cite{gobert05,vidmar15} (compare the sketch in \fref{fig:setup}). 
From the two boundaries,  two ``light cones'' emerge, consisting of
particles outside and holes inside the block. Both particles and holes have a maximum speed of $v_m
= 2 \, J_x a$. 
Consequently, the time $t_2 := B/4 J_x$ is the earliest possible time at which the melting arrives at the
center, such that the density drops below one on all sites.
Thus, $t_2$ marks the point in time at which
density profiles obtained from blocks start to differ quantitatively from those of domain walls, defining the third time regime.
In the case of a ballistic expansion realized for $J_y \ll J_x$, the density in the center drops
strongly at $t_2$ and we can clearly identify two outgoing ``jets'' as two separating maxima in the density
profiles; see \fref{fig:cylprofilesr}(a). 
To be clear, the expectation for the nature of mass transport in a nonintegrable model such as coupled
systems of 1D hard-core bosons is diffusion, sustained by numerical studies \cite{steinigeweg14}.
However, in the sudden expansion, the whole cloud expands and it is conceivable that the 
expansion appears to be ballistic because the cloud becomes dilute too fast, resulting in mean-free paths
being on the order of or larger than the cloud size at any time \cite{vidmar13}.

On the other hand, for larger $J_y$ the block in the center does not split at $t_2$,
but  a region with a high density (``core'') remains in the center. 
The high-density core is clearly established already at  intermediate $J_y/J_x= 0.5$, where it still
expands slowly. For
larger $J_y$, the spreading of this core is continuously suppressed.

\subsection{Integrated current}
\begin{figure}[tb]
\includegraphics[width=\columnwidth]{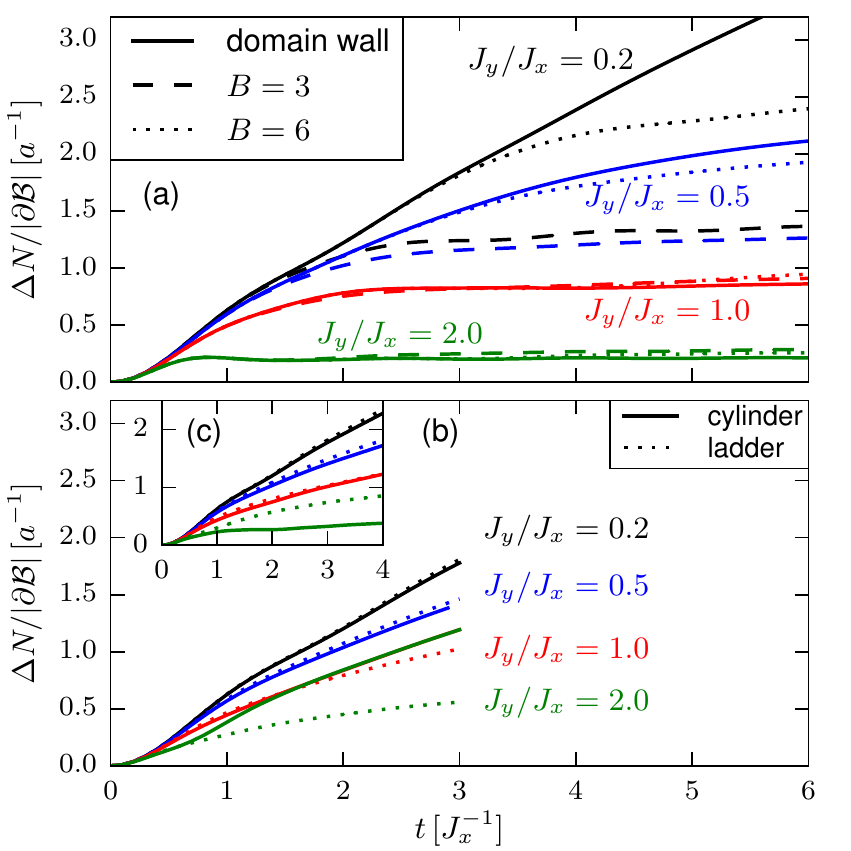}
\caption{(Color online) Comparison of $\Delta N / |\partial \mc{B}|$ (a) on a two-leg ladder for the expansion from central
	blocks (dashed/dotted lines, $B \times 2$ bosons) versus the melting of a domain wall (solid lines). 
	The lower panels compare domain walls on ladders (dashed lines) to domain walls on cylinders (solid lines) for (b) $L_y=4$  and (c)
	$L_y=3$. 
	The curves for  the $L_y=4$ cylinder in (b) with $J_y/J_x=1,2$ are nearly on top of each other
	for $t \gtrsim 1.5 \,J_x^{-1}$.
	}
\label{fig:DeltaN}
\end{figure}

In order to investigate the different time regimes further,
we consider the number of bosons $\Delta N(t)$ that at a time $t$ have left the block
 $\mc{B}$ where they were initialized.
This is equivalent to the particle current  
$j^x_{i_x} = \mathrm{i} J_x \sum_{i_y} \langle \ahatdag_{i_x+1,i_y} \ahat_{i_x,i_y} -\ahatdag_{i_x,i_y} \ahat_{i_x+1,i_y} \rangle $
integrated over time and along the boundary $\partial \mathcal{B}$ of the block,
\begin{equation}
	\Delta N (t) = \sum_{i \notin \mc{B}} n_{i_x,i_y}(t) 
	= \int_0^t {\rm d}s\, \left[j^x_{b_r}(s) - j^x_{b_l}(s) \right].
	\label{eq:DeltaN}
\end{equation}
Here $b_r$ and $b_l$ denote the right and left indices $i_x$ of the boundary of the initially
centered block $\mc{B}$. 
We compare $\Delta N$  for the expansion on a two-leg ladder starting from either central blocks or
domain walls in \fref{fig:DeltaN}(a).
To this end we normalize $\Delta N$ by the boundary length $|\partial \mc{B}|$, which is simply 
$2 L_y \, a$ for the central blocks and $L_y \, a$ for the domain walls. 

For short times $ t \lesssim 0.5 \, J_x^{-1}$ (i.e., $t\lesssim t_1$, see the above), all curves in \fref{fig:DeltaN} are independent of $J_y$.
For the quantity $\Delta N$, the first deviations between domain walls and cylinders do not occur at
$t_2$ but at $2t_2 = B/2 J_x $,
which is exactly the time the fastest holes need to travel once completely through the block:
By definition, $\Delta N$ is not sensitive to the density inside the initial block.
For the expansion of central blocks, particle conservation gives a strict bound 
$\Delta N/|\partial\mc{B}| \leq B/2a$, in which case all the bosons have
left the initial block. These bounds (equal to $1.5\,a^{-1}$ and $3\, a^{-1}$ for $B=3 $ and $B=6$, respectively)
are approached in the long-time limit of the ballistic expansion for small $J_y/J_x =0.2 $, which
for $B=6$, however, happens beyond the times reached in our simulations.
For the domain walls, $\Delta N $ is not bounded (as long as the melting does not reach the boundary
of the system) and grows for small $J_y/J_x$ as $\Delta N\propto t$ linearly in time, which, via  \eref{eq:DeltaN}, corresponds to a nondecaying current $j^x$.
On the other hand, $\Delta N$ gets almost constant for large $J_y/J_x$ for both the domain walls and
the blocks. This indicates that the expansion is strongly suppressed on the two-leg ladder, with a high-density core remaining in the center.
We speculate that the regime in which $\Delta N$ increases only very slowly is indicative of diffusive dynamics, by similarity with \cite{schneider12}. 

\subsection{Propagating modes: Limit of large $J_y \gg J_x$}
\label{sec:PropagationModes}
In order to qualitatively understand the suppression of the expansion for certain geometries and specific values of $L_y$,
it is very instructive to consider the limit of large $J_y \gg J_x$. 
We discuss this limit in more detail in Appendix \ref{sec:detailedlargeJy}, while here, we provide only the general idea
and discuss the results.
The Hamiltonian \eref{eq:ham} can be split up into the hopping on rungs 
(we denote sites with the same index $i_x$ as a ``rung'' for both ladders and cylinders),
denoted by $H^y$ proportional to $J_y$, and the hopping terms in the $x$ direction proportional to $J_x$, collected in $H^x$.
Our analysis is based on a diagonalization of $H^y$, which is a block-diagonal product of
terms operating on single rungs.
We view the eigenstates of single rungs as ``modes,'' which can be delocalized by $H^x$. 
Since a coherent movement of multiple bosons is a higher-order process of $H^x$ and thus generally
suppressed for large $J_y/J_x$, we focus on modes with a single particle on a rung.
We then look for modes which are candidates for a propagation at finite $k_x$.
Importantly, the kinetic energy $E_x \propto J_x$ cannot compensate for a finite $E_y \propto J_y$
for $J_y \gg J_x$.
Since we initialize the system in states with zero total energy,
energy conservation allows only modes with $E_y = 0$ to contribute to the expansion in first-order processes in $J_x/J_y$ in time.
In general, one could also imagine to create pairs of two separate bosons with exactly opposite
$E_y$, summing up to 0.
Yet  $H^x$ cannot create such pairs (see Appendix \ref{sec:detailedlargeJy} for details).

For smaller $J_y$, the scaling argument of the energy conservation
does not hold and additional modes (beginning with those of small energy $E_y$) can be used for the
propagation in the the $x$ direction; ultimately, for $J_y \ll J_x$ any mode contributes to the expansion already at short times.
We note that modes with strictly $E_y= 0 $ are either present or absent at any value of $J_y/ J_x$.

Such propagating single-boson modes with $E_y=0$ do {\em not} exist on a two-leg ladder:
There are, apart from  the  empty and filled rung, only two states with large energies $E_y=\pm J_y$. 
We argue that precisely this lack of modes with $E_y=0$ leads to the suppression of the expansion with increasing $J_y/J_x$.
It is manifest in \fref{fig:DeltaN}(a) by the fact that $\Delta N$ gets almost constant.
Thus, we can view the expansion to be inhibited by the existence of heavy objects (particles of a large effective mass)
that can propagate only via higher-order processes. This is  similar to the reduction of expansion velocities due to
doublons in the strongly interacting regime of the 1D Bose-Hubbard model \cite{ronzheimer13,vidmar13,boschi14,sorg14,weiss}.
Another  effect with very similar physics is self-trapping (see, e.g., \cite{trombettoni01,hennig13,jreissaty13}).

\begin{figure*}[tb]
\includegraphics[width=\textwidth]{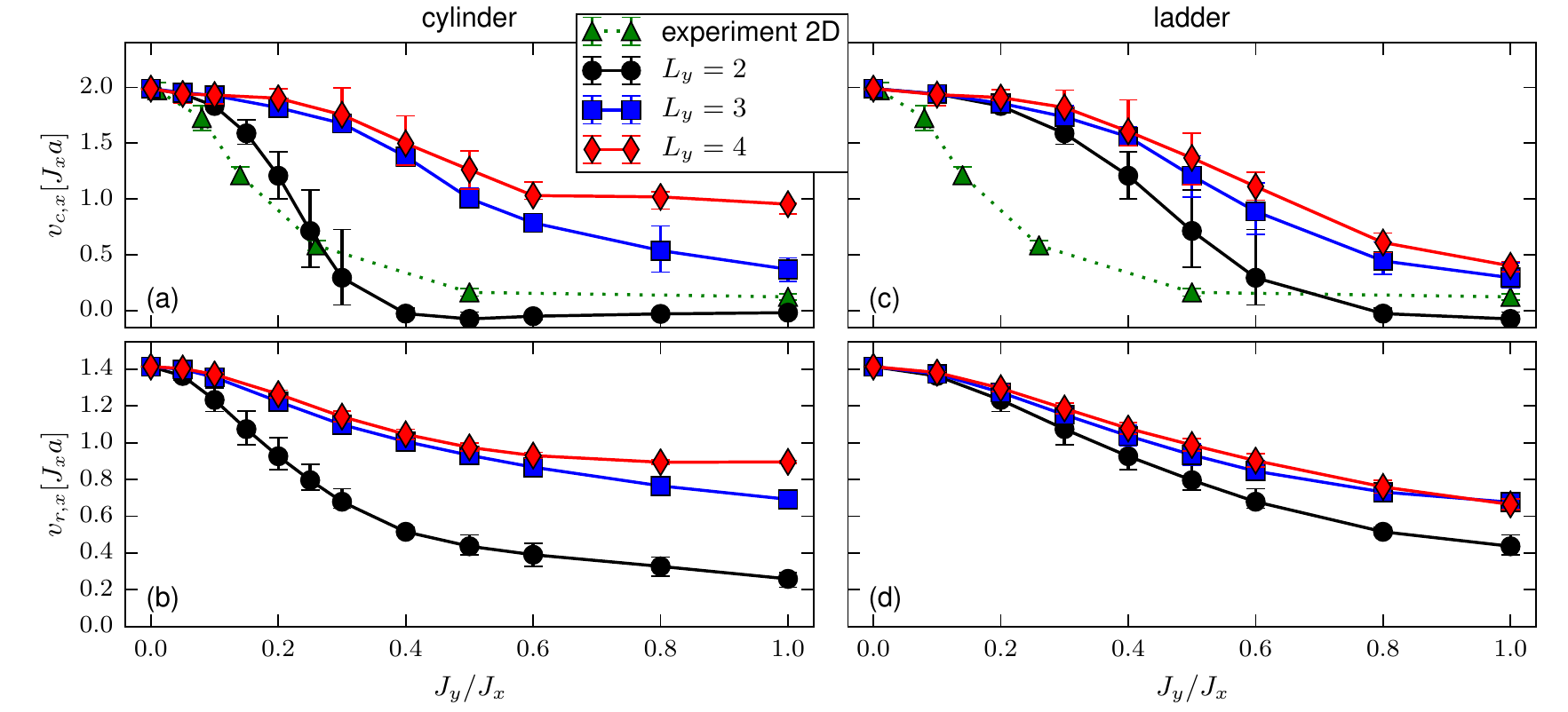}
\caption{(Color online) (a),(c) Core expansion velocities $v_{c,x}$ 
	and (b),(d) radial velocities $v_{r,x}$ versus $J_y$ for the expansion of a $6\times L_y$ block.
	The left panels (a) and (b) are obtained on $L_y=2,3,4$ cylinders; the right
	panels (c),(d) are obtained on $L_y=2,3,4$ ladders.
	The green triangles taken from Ref.~\cite{ronzheimer13} show the results of the experiments for
	the fully 2D expansion corresponding to the setup of \sref{sec:2Dexpansion}.
}
\label{fig:cylladvcvr}
\end{figure*}

Whether propagating modes with $E_y=0$ exist or not depends not only on $L_y$ but also on the boundary conditions in
the $y$ direction.  This can serve as a test for our reasoning.
For $L_y=4$, we find modes with $E_y=0$ on a cylinder but not on a ladder (see
Appendix~\ref{sec:detailedlargeJy}).
We compare $\Delta N$ for these two geometries directly in Figs.~\ref{fig:DeltaN}(b) and \ref{fig:DeltaN}(c). 
For small $J_y/J_x = 0.2$, the additional coupling
of the cylinders compared to the ladders has 
(at least on the time scales accessible to us) nearly no influence.
Yet, for large $J_y/J_x$, we find not only a quantitative but even a qualitative difference:
For the $L_y = 4 $ cylinders, $ \Delta N$ increases linearly in time, irrespective of how large $J_y/J_x$ is.
Moreover, the slope is (at $t \gtrsim 1.5 \,J_x^{-1}$) roughly the same for all $J_y/J_x \gtrsim 0.5$
and does almost not decrease with time.
Using \eref{eq:DeltaN}, we can relate this to the presence of  a non-decaying current, which we
explain in terms of an enhanced occupation at momenta compatible  with $E_y = 0$.
In contrast, on the four-leg ladder there are no propagating modes with $E_y = 0$; thus, we expect no
linear increase of $\Delta N$.
Indeed, we find that the currents---i.e., the slopes of $\Delta N$ in \fref{fig:DeltaN}(c)---on the four-leg ladder decay in time.
Yet, the decay is not as extreme as for the two-leg ladder,
which we explain by the existence of modes with lower energies $E_y>0$ than on the two-leg ladder.
For $L_y= 3$, it is exactly the other way around: There are modes with $E_y = 0$ on the ladder but
not on the cylinder. In agreement with this,  \fref{fig:DeltaN}(c) shows that the
expansion on a three-leg ladder is faster than on an $L_y=3$ cylinder for large $J_y/J_x = 2$.

\subsection{Expansion velocities}
\label{sec:velo}
Figure \ref{fig:cylladvcvr} shows the radial and core velocities for the
expansion of blocks on cylinders and ladders.
We note that, while $v_c$ and $v_r$ are nearly independent of $J_y/J_x $ in the range $ J_y/J_x= 0.6,
\dots, 1$ for the $L_y=4$ cylinder [Figs.~\ref{fig:cylladvcvr}(a) and \ref{fig:cylladvcvr}(b)],
 the values $r_c(t)$ and $\tilde{R}(t)$ themselves actually do decrease when $J_y/J_x$ is tuned
from 0.6 to 1 (see Figs.~\ref{fig:cylR} and \ref{fig:cylrc} in the Appendixes),
due to different short-time dynamics.
Further, for the accessible times ($t_m =3 \,J_x^{-1}$ for $L_y=4$), 
the density profile outside the original block is still completely equivalent to  the domain-wall melting.
Nevertheless, $r_c(t)$, by definition, is also sensitive to the maximum value in the center of the
block, and $\tilde R(t)$ is sensitive to the densities at all positions.
Thus, the  velocities shown in \fref{fig:cylladvcvr} contain valuable and complementary information.

The two-leg ladder (for which the expansion velocity has been studied  in Ref.~\cite{vidmar13}) shows a behavior similar to the
experimental data for 2D expansions \cite{ronzheimer13}, namely that the core velocity $v_c$ drops down to
zero with increasing $J_y/J_x$.
However, by comparing different $L_y$, we find a trend towards a faster expansion when $L_y$ is increased at fixed $J_y/J_x$.
This trend is in contrast to the naive expectation that wider cylinders should mimic the
1D-to-2D crossover better. 
In other words, it demonstrates that the two-leg ladder does not capture all the relevant
physics of the expansion in all directions in the 1D-to-2D crossover, although it shows  the same qualitative dependence of velocities on $J_y/J_x$ as the  2D system studied experimentally \cite{ronzheimer13}.
However, we understand this from our considerations of the limit $J_y\gg J_x$ in
 Sec.~\ref{sec:PropagationModes}:
On the $L_y=4$ cylinder and the $L_y=3$ ladder, there exist  $E_y=0$ modes, and thus a preferred occupation of these  
propagating modes with nonzero $k_y$ is possible. Moreover, in those other cases in which there are no modes with strictly
$E_y = 0$, there are at least modes with lower $|E_y|<J_y$.

\subsection{Momentum distribution function}
\begin{figure}[tb]
\includegraphics[width=\columnwidth]{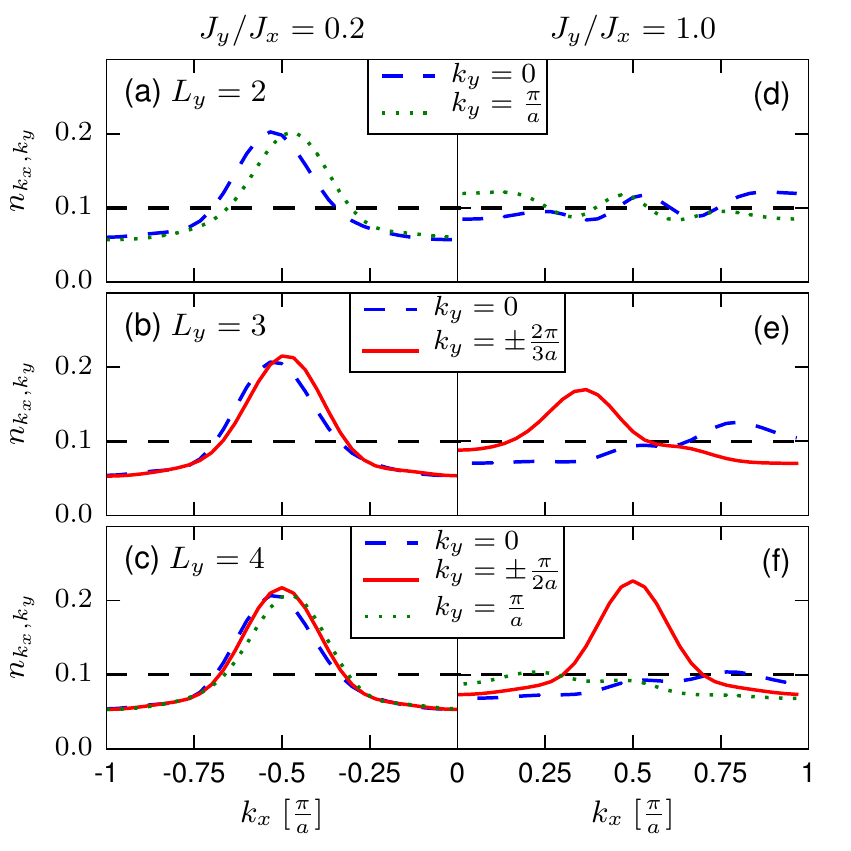}
\caption{(Color online) Momentum distribution function $n_{k_x,k_y}$ (dimensionless) for cylinders with (a),(d) $L_y=2$, (b),(e) $L_y=3$
	 and (c),(f) $L_y=4$, starting from a $6\times L_y$ cluster. Data are shown for time $t =2.0 \,J_x^{-1}$ and (a)--(c)
	$J_y/J_x = 0.2$ and (d)--(f) $J_y/J_x= 1.0$ (Note that we have a symmetry $n_{-k_x, k_y} = n_{k_x, k_y} = n_{k_x, -k_y}$).
	The black dashed lines indicate the flat initial distribution at $ t = 0$.
} 
\label{fig:cylNk}
\end{figure}

The momentum distribution $n_{k_x,k_y}$ on cylinders starting from $6\times L_y$ blocks and at fixed time $t = 2.0 \,J_x^{-1}$  is shown in \fref{fig:cylNk}.
At small $J_y/J_x = 0.2$, we observe a bunching of particles at  the  $k_x = \pm \frac{\pi}{2a}$ modes independent of
$k_y$, similar to  the fully 2D expansion at the same value of $J_y/J_x$ shown in \fref{fig:2Dmomentum}.

For $J_y=J_x$ and on the $L_y = 3$ cylinder, the energy $E_y(k_y=\pm \frac{2\pi}{3a}) = J_y$ can be compensated
by kinetic energy $E_x = - 2 J_x \cos(k_x a)$ in the $x$ direction; compare \eref{eq:EnergyCondensates}.
Indeed, we find a bunching of particles at those momenta in \fref{fig:cylNk}(e). 
The $E_y(k_y=0)=-2J_y$ and $E_y(k_y= \frac{\pi}{a}) = 2 J_y$ mode would
yield $k_x = \frac{\pi}{a}$ and $k_y = 0$, yet we find a slightly higher weight at smaller $k_x$ in
\fref{fig:cylNk}(e).
However, we note that all these peaks for $J_y = J_x$ in Figs.~\ref{fig:cylNk}(d) and \ref{fig:cylNk}(e) are not as high as their counterparts for $J_y/J_x = 0.2$.
As we have discussed above and in Appendix \ref{sec:detailedlargeJy}, there are no modes with
$E_y=0$ for $L_y=2,3$ on cylinders; hence, the maxima in $n_{k_x,k_y}$ are generally suppressed as we go from small to large $J_y/J_x$ 
for $L_y=2,3$.

On the $L_y=4$ cylinder, we find a bunching of particles at $(k_x, k_y) = (\frac{\pi}{2a}, \frac{\pi}{2a})$
with roughly the same weight for all $J_y$; compare Figs.~\ref{fig:cylNk}(c) and \ref{fig:cylNk}(f).
This is in agreement with our considerations of \sref{sec:PropagationModes}, since the modes with $k_y
= \frac{\pi}{2a}$ have $E_y = 0$. The $k_y=0, \frac{\pi}{a}$ modes are suppressed, similar to  the case of $L_y=2,3$.

The question of whether the bunching of particles at certain quasimomenta (that requires the existence of propagating modes
with energies compatible with those quasimomenta) 
will lead to a true dynamical quasicondensation at finite momenta can best be addressed using the 
domain walls as initial stats.
Here, we are guided by the behavior of 1D hard-core bosons: In the sudden expansion \cite{rigol04,vidmar13}, the dynamical quasicondensation
is a transient phenomenon, hence the occupation at $k=\pm \frac{\pi}{2a}$ first increases and then slowly decreases as dynamical
fermionization sets in \cite{rigol05a,minguzzi05,vidmar13}. The crossover between these two regimes---the formation and the decay of quasicondensates---is
given by $t_2 \propto B$ (see also the discussion in \cite{vidmar15}).
For the domain-wall melting, the quasicondensates are continuously fed with particles with identical properties due to the presence of an infinite reservoir and  thus the quasicondensation peaks
in $n_k$ never decay but keep increasing. 
 
\begin{figure}[tb]
\includegraphics[width=\columnwidth]{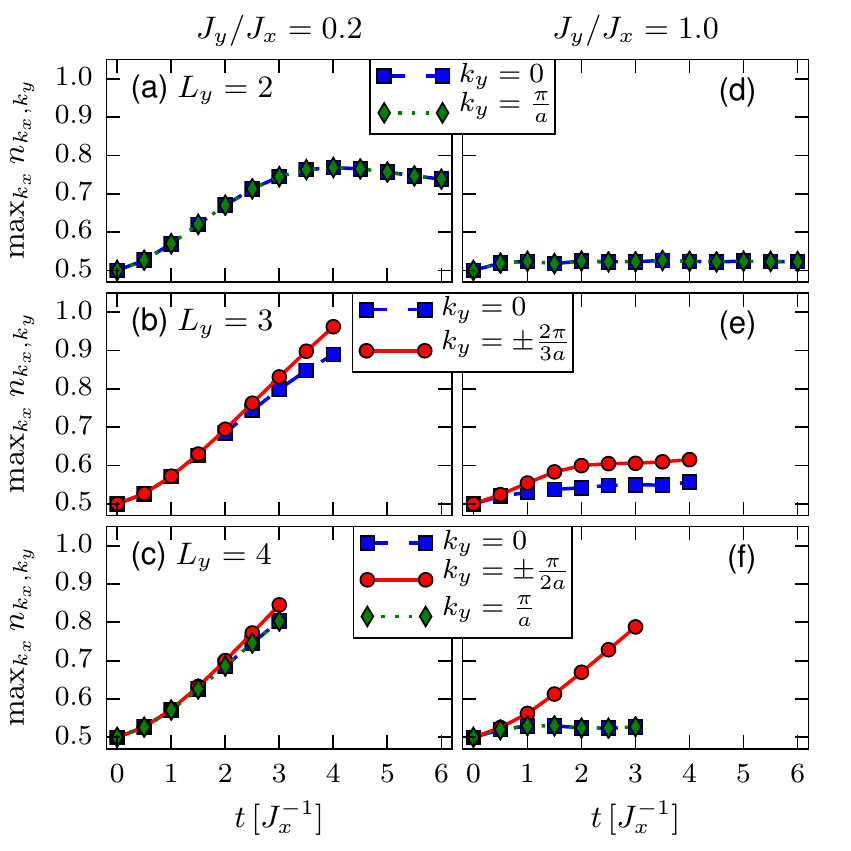}
\caption{(Color online) Time evolution of the peak heights in the momentum distribution function for cylinders with 
	(a),(d) $L_y=2$, (b),(e) $L_y=3$, and (c),(f) $L_y=4$, starting from a domain wall. Data are shown for (a)--(c)
	$J_y/J_x = 0.2$ and (d)--(f) $J_y/J_x= 1.0$.
} 
\label{fig:cylmaxNkdw}
\end{figure}
Figure~\ref{fig:cylmaxNkdw} shows the time dependence of the occupation 
at the maximum of $n_{k_x,k_y}$ for the domain-wall melting on $L_y=2,3,4$ cylinders for (a)--(c)
$J_y/J_x=0.2$ and (d)--(f) $J_y/J_x=1$.
For $J_y/J_x=0.2$ and the accessible time windows of the $L_y = 3, 4$ cylinders, the occupation
indeed increases monotonically in time. 
On the $L_y=2$ cylinder in \fref{fig:cylmaxNkdw}(a), the maximum initially increases similar as
for $L_y=3, 4 $, yet for times $t \gtrsim 3 J_x^{-1}$ it saturates and even decreases,
which suggests that no condensation sets in. Note that the time scale at which the saturation
happens is quite large, as it is set by $J_y^{-1}$. 
This suggests that there is no condensation even for very small $J_y>0$ on the $L_y=2$ cylinder.

The behavior for $J_y/J_x=1$ is quite different. In almost all cases, the occupation at the maximum quickly saturates, which suggests that
no condensation sets in. This observation is consistent with the absence of fast propagating modes on the 
$L_y=2,3$ cylinders.
Among the data sets shown in Fig.~\ref{fig:cylmaxNkdw}(d)--\ref{fig:cylmaxNkdw}(f), there is one exception, namely the peak at
$(k_x,k_y)=(\frac{\pi}{2a},\pm \frac{\pi}{2a})$
on the four-leg cylinder, which monotonically increases without a trend towards saturation.
This case is thus the most promising candidate for a condensation at $J_y = J_x$.

\subsection{Occupation of lowest natural orbital}
\begin{figure}[tb]
\includegraphics[width=\columnwidth]{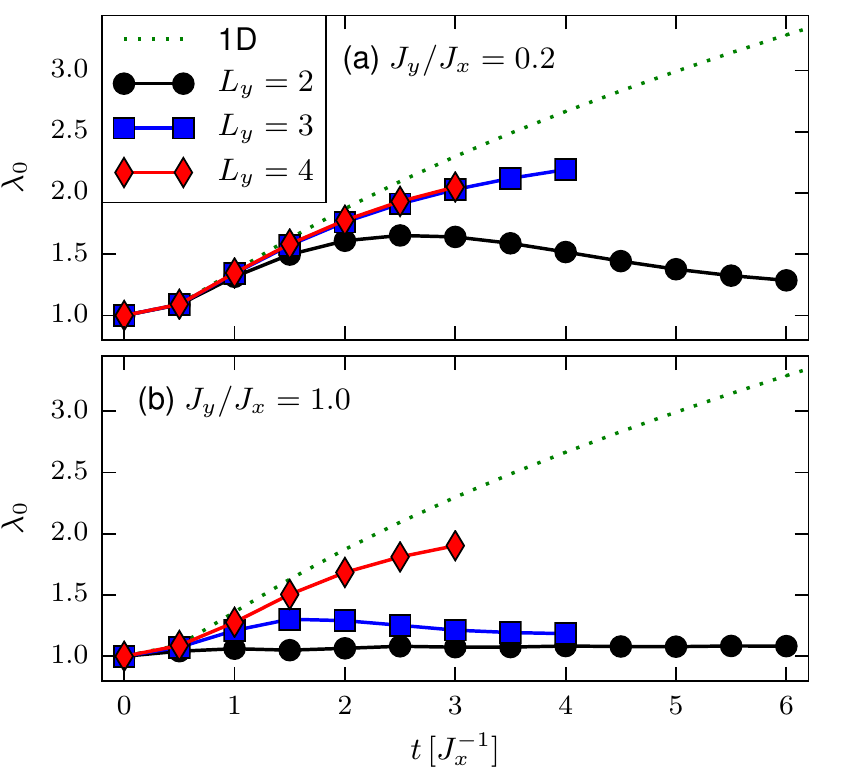}
\caption{(Color online) Time evolution of the occupation of the largest eigenvalue $\lambda_0$ (dimensionless) of the 
	one particle density matrix for cylinders with (a) $J_y/J_x = 0.2$ and (b) $J_y/J_x= 1.0$,
	starting from a domain wall.
	The dotted green lines show the results of an 1D chain ($J_y=0$) for comparison.} 
\label{fig:cylmaxNOdw}
\end{figure}

To investigate the question of condensation in more detail, we look at the maximum occupation $\lambda_0$ of the natural
orbitals \cite{penroseonsager56}.
The natural orbitals are effective single particle states defined as the eigenstates of the one particle density matrix 
$ \braket{\ahatdag_{\mathbf{i}}\ahat_{\mathbf{j}}}$.
The corresponding eigenvalues sum up to the number of particles and can be interpreted as the occupations of the natural orbitals.
A true condensate requires that $\lambda_0$ becomes macroscopically large.

The largest occupation $\lambda_0$ for the domain-wall melting of cylinders is shown in \fref{fig:cylmaxNOdw}.
In the 1D case, indicated by the green dotted line, the occupation grows, for large times, as $\lambda_0 \approx 1.38 \sqrt{t}$ \cite{rigol04}.
For $L_y = 2$ we find two degenerate natural orbitals with occupation $\lambda_0$.
For $J_y/J_x=0.2$,  
we find an initial growth for all $L_y = 2,3,4$, but for $L_y=2 $, the occupation saturates and even
decreases for large times $t \gtrsim 3 J_x^{-1}$, similar as for the peaks in the momentum distribution function. 
In fact, the peaks in the momentum distribution are directly related to the natural orbitals with
the largest occupation: For $L_y = 2$ there are two degenerate natural orbitals with maximal
occupation with $k_y = 0$ and $k_y = \frac{\pi}{a}$, and their Fourier transformation is
peaked slightly above (below) $k_x = \frac{\pi}{2a}$ for $k_y = 0$ ($k_y=\frac{\pi}{a}$).
Similarly, for $L_y = 3$ ($L_y=4$) there are two natural orbitals with maximal occupation with $k_y =
\pm\frac{2\pi}{3a}$ ($k_y=\pm \frac{\pi}{2a}$) and one (two) with slightly lower occupation with
$k_y= 0$ ($k_y=0, \frac{\pi}{a}$), leading to the peak structure of Figs.~\ref{fig:cylNk}(b) and \ref{fig:cylNk}(c) (with peaks only at $k_x > 0$ for domain-wall initial states).

For $J_y = J_x$, shown in \fref{fig:cylmaxNOdw}(b), $\lambda_0$ saturates and even decreases for the cylinders of
width $L_y = 2,3$,
but keeps growing monotonically for $L_y = 4$ (at least on the time scale accessible to us), in accordance with Figs.~\ref{fig:cylNk}(f) and \ref{fig:cylmaxNkdw}(f).
For $L_y = 4$, we find only two (degenerate) natural orbitals with $k_y = \pm \frac{\pi}{2a}$ with
peaks at $k_x = \frac{\pi}{2a}$.
Yet the maximal occupation $\lambda_0 $ is significantly smaller than in the 1D case and seems to saturate at larger times.

It is instructive to compare $\lambda_0$ to the number of particles in the expanding cloud $\Delta
N$ shown in \fref{fig:DeltaN}, defining a condensate fraction $\lambda_0/\Delta N$.
$\Delta N$ increases linearly in time in 1D; hence, the condensate fraction goes to zero with
$1/\sqrt{t}$, consistent with the absence of true long-range order. 
In the case of cylinders, we never observe a saturation of $\lambda_0/\Delta N$ to a constant
nonzero value, but it keeps decreasing as a function of time. 
Therefore, a true condensation is not supported by the existing data on any cylinder.
Yet the survival of a quasicondensation on the cylinders is consistent with our data.

\section{Summary}
\label{sec:sum}

Motivated by recent experiments with ultracold bosons in an optical lattice \cite{ronzheimer13,vidmar15}, we simulated the sudden expansion of up to $4\times4$ hard-core bosons in a 2D lattice.
In the limit $J_x \gg J_y$, we find a fast expansion (at least on the time scale
accessible to us), similar to the 1D case.
When $J_y$ is tuned to the isotropic limit $J_x = J_y$, some fraction of the particles remains as a high-density core in the
center and a spherically symmetric shape emerges. This trend is compatible with the observations made in the 
experiment of Ref.~\cite{ronzheimer13}.
Unfortunately, our results for the 2D expansion are dominated by surface effects due to
the small boson numbers as, in fact, in our simulations we have more particles at the boundary of the initial block than in the
bulk. This prevents us from analyzing the core expansion velocity \cite{ronzheimer13}, yet the radial
velocities $v_{r,x}$ decrease monotonically with the block size $B$  at any fixed $J_y/J_x$.
We observe a bunching in the momentum distribution function at quasimomenta compatible with
energy conservation.
This bunching could signal a dynamical condensation at finite quasimomenta as in the 1D case, where this dynamical quasicondensation \cite{rigol04}
has recently been observed in an experiment \cite{vidmar15}.
Although we cannot ultimately clarify the question of dynamical condensation in 2D with our small clusters, we believe that the bunching of particles at certain finite momenta 
in the
2D expansion $J_y \approx J_x$ stems from surface effects.

In order to investigate the dimensional crossover further, we studied the expansion on long cylinders and
ladders with up to $L_y=4$ legs.
Correlations between the particles in different legs, which lead to a $J_y$ dependence, built up on a very short timescale of about one tunneling time in the longitudinal $x$ direction.
Up to a time $t_2$ that is proportional to the linear dimension of the initial block, 
the expansion of blocks, restricted to either the left or right half of the system, is 
identical to the domain-wall melting.
On two-leg ladders, the density in the central region becomes very weakly time dependent and almost stationary for $J_y/J_x \gtrsim 1$, even for the domain walls.
This is  reflected by a vanishing or even slightly negative core
velocity, similar to the observations made in experiments \cite{schneider12,ronzheimer13}.
By considering the limit $J_y \gg J_x$, we argue that this suppressed expansion on the two-leg ladder for large $J_y/J_x$ 
stems from the fact that there are no modes with $E_y = 0$ on single rungs.
For cylinders and ladders with larger $L_y\in \set{3,4}$, we generically find a faster expansion with higher
velocities than in the $L_y=2$ case. 
Additionally, there is a dependence of expansion velocities on the boundary conditions in the $y$ direction.
For instance, the expansion on $L_y=4$ cylinder is faster than on a four-leg ladder.
In agreement with our considerations of the limit $J_y\gg J_x$, this is accompanied by a bunching at
preferred momenta $k_y = \pm \frac{\pi}{2a}$ and $k_x = \pm \frac{\pi}{2a}$ and an increasing
occupation of natural orbitals.
Yet our data does not support a true condensation on any cylinder.

Finally, we state the interesting  question whether the expansion velocities on cylinders or ladders
will ever show the same dependence on $J_y/J_x$ as the width $L_y$ increases compared to the
expansion of a 2D block.
The obvious difference is  that we fill the cylinders and ladders completely in the
$y$ direction. 
Due to symmetry, the expansion on cylinders is restricted to be along the $x$ direction and, as such,
closer to the  1D case, at least for small $L_y$.
There can thus be two scenarios:
Either,  even for $L_y \rightarrow \infty$, the velocities of the cylinders 
might well be above the experimental results or, as $L_y$ increases beyond $L_y=4$, the velocities
at a fixed $J_y/J_x$ will depend nonmonotonically on $L_y$.

Further insight into these questions, i.e., the dependence on $L_y$ or the question of dynamical condensation at finite momenta in dimensions higher than one, could be gained from future experiments with access to measuring the radius. 
This could be accomplished using single-site resolution techniques; see \cite{preiss15,fukuhara13,fukuhara13a}
for work in this direction.

{\it Acknowledgments.}
We are indebted to M.~Rigol for valuable comments on a previous version of the manuscript.
We acknowledge useful discussions with I.~Bloch, L.~Pollet, M.~Rigol, U.~Schneider, and L.~Vidmar.
F.H.-M.~was supported by DFG (Deutsche Forschungsgemeinschaft) Research Unit  FOR 1807 through Grant No.~HE~5242/3-1.
This work was also supported in part by National Science Foundation Grant No.~PHYS-1066293 and the hospitality of the Aspen Center for Physics.


\renewcommand{\thetable}{A\arabic{table}}
\renewcommand{\theequation}{A\arabic{equation}}
\renewcommand{\thesection}{A\arabic{section}}

\appendix
\section{Extraction of core and radial velocity}
\label{sec:extractvcvr}

\begin{figure}[!t]
\includegraphics[width=\columnwidth]{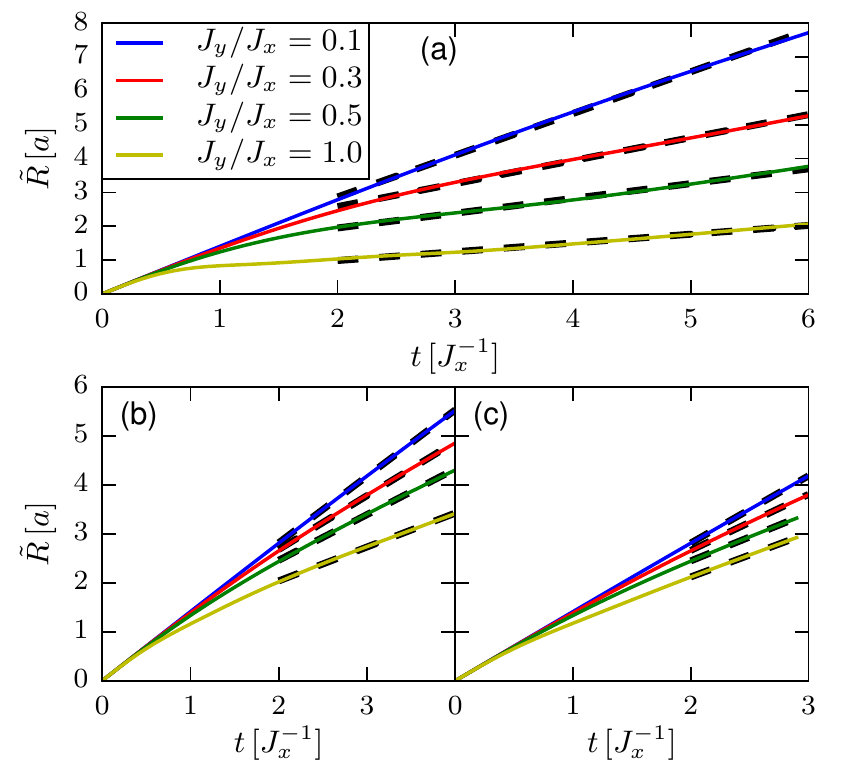}
\caption{(Color online) Reduced radius $\tilde{R}(t)$ for cylinders with (a) $L_y=2$, (b) $L_y=3$, and 
(c) $L_y=4$, starting from a $6\times L_y$ cluster. 
The thick dashed lines show the linear fits used to extract the radial velocities $v_r$, which are shown in
\fref{fig:cylladvcvr}(b).
}
\label{fig:cylR}
\end{figure}

\begin{figure}[tb]
\includegraphics[width=\columnwidth]{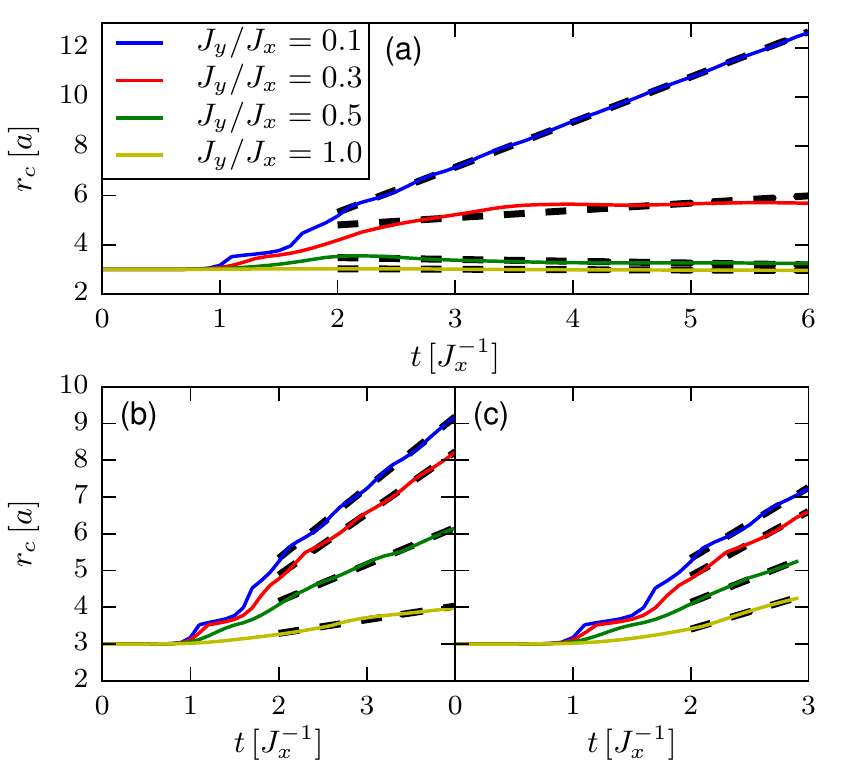}
\caption{(Color online) Core radius $r_c(t)$ for cylinders with (a) $L_y=2$, (b) $L_y=3$, and 
(c) $L_y=4$, starting from a $6\times L_y$ cluster. 
The thick dashed lines show the linear fits used to extract the core velocities $v_c$, which are shown in 
\fref{fig:cylladvcvr}(a).}
\label{fig:cylrc}
\end{figure}

Both velocities $v_r = \frac{\partial \tilde{R}(t)}{\partial t} $ and 
$v_c = \frac{\partial r_c(t)}{\partial t}$ are
time derivatives of quantities which are not strictly linear in time.
Thus, both $v_r$ and $v_c$ themselves are time dependent. 
Figure \ref{fig:cylR} shows the time dependence of the reduced radius $\tilde{R}(t)$,
while \fref{fig:cylrc} shows the core radius $r_c(t)$.
In the ideal case we would expect them to get constant in the long-time limit. Unfortunately, our calculations are limited to finite times  $t_m=6 \,J_x^{-1}$ for the two-leg ladder, 
$t_m \approx 4 \,J_x^{-1}$ for $L_y=3$ cylinders/ladders, $t_m \approx 3 \,J_x^{-1}$ for $L_y = 4$
cylinders/ladders, and just $t_m \approx 1.5 \,J_x^{-1}$ for the 2D lattice.

The reduced radii all start as $ \tilde{R} (t) = \sqrt{2}t J_x a $ on very short time scales $ t \lesssim 0.5 \,J_x^{-1}$.
This is clear as we are initially confined to the hopping in the $x$ direction, independently of $J_y$.
For very small $J_y \ll J_x$, the reduced radius remains linear in time with the velocity $v_r =
\sqrt{2}J_xa$ at all
times, as expected for a ballistic expansion from an initial state with a flat quasimomentum distribution function \cite{langer12,ronzheimer13}.
A $J_y$ dependence may show up on a timescale $ t \propto J_y^{-1}$.
For larger $J_y$ the slope $v_r$ reduces at intermediate times (in the time range where we can
observe it) but increases again for large $t J_x$. The latter can be understood as follows:
The outermost parts have the strongest contribution to the sum in \eref{eq:R}, and naturally these
outer parts have the highest velocity $2\, J_x a$ (and also reached a low density such that they are dilute
and thus do not see  each other any more).
Assuming a fraction $p$ of the particles to expand with $v$ and the rest $(1-p)$ to form an
inert time-independent block in the center (see also the argument given in \cite{sorg14}), a
straightforward calculation shows that $\tilde{R}(t) \approx \sqrt{p}\, v\, t$ at large times.
This is also the reason why $\tilde{R}(t)$ does {\em not} settle to a constant value on the two-leg
ladder even for large $J_y$, 
although the core in the center barely melts and $\Delta N$ becomes only weakly time dependent: There is always a nonzero fraction of particles 
which go out  from the center.

We extract the time-independent expansion velocities $v_r$ shown in Figs.~\ref{fig:2Dvr} and
\ref{fig:cylladvcvr} by a linear fit $\tilde{R}(t) = v_r \cdot t + \mathrm{const}$ 
in the time interval $ 2.0 \,J_x^{-1}\leq t \leq t_m$, where $t_m$ is the maximum time reached in the simulations; see the above.
For the 2D lattice, we reach only $t_m = 1.5 \,J_x^{-1}$; thus, we fit only in the interval 
$1.0 \,J_x^{-1} \leq t \leq 1.5 \,J_x^{-1}$ in this case.
In \fref{fig:cylladvcvr} we show error bars resulting from similar fits but using only the first or the second
half of the time interval.

In the  time regime $ 0 < t < t_2 $, the core radius is constant, although the
cloud already expands: From both edges, the block melts, but
the location of the half-maximum density does not move due to particle-hole symmetry.
Just when the first holes arrive in the center of the block, the global maximum decreases and $r_c$,
the half width at half maximum, begins to increase.
It then exhibits strong initial oscillations. The latter stem, on the one hand, 
from the discreteness of the particles' coordinates on the lattice, which is only partly cured by the linear splines used to
extract $r_c$. On the other hand, the melting of domain walls in 1D happens in quantized
``charges,'' which lead to well-defined structures in the density profile \cite{gobert05,hunyadi04,eisler13}.
Those oscillations prevent us from extracting the core velocity for the 2D lattice,
where they are too strong at the times  reached in the simulations. 
Yet  it seems reasonable to extract $v_c$ for the cylinders and ladders by linear fits $r_c(t) = v_c \cdot t + \mathrm{const}$ in
the same way as for $v_r$. While it works quite well for the ballistic expansion at $J_y \ll J_x$  and
quite large $J_y \gtrsim J_x$, $r_c(t)$ still exhibits a stronger time dependence for intermediate
$J_y$, e.g., $J_y\approx 0.3 J_x$ on the $L_y=2$ cylinder. 
In the latter case, some of the bosons expand initially during the domain-wall melting and thus the block 
and $r_c$ grow, yet then the expansion is slowed down and the extension of the high-density block measured by $r_c$ 
becomes weakly time dependent.

\section{Limit of large $J_y \gg J_x$}
\label{sec:detailedlargeJy}

We split the Hamiltonian (\ref{eq:ham}) into two parts according to  $H = \sum_{i_x} ( H^y_{i_x} + H^x_{i_x, i_x+1} ) $, where
$H^y_{i_x} = - J_y \sum_{i_y}(\ahatdag_{i_x,i_y} \ahat_{i_x,i_y+1} +h.c.) $ collects the hopping terms within the rung $i_x$ 
and $H^x_{i_x,i_x+1}$ collects the hopping terms between neighboring rungs.

\subsection{Two-leg ladder}

In the following we give an explicit expression for  $H^x_{i_x,i_x+1}$  on a two-leg ladder
in terms of the eigenstates of $H^y_{i_x}$ and $H^y_{i_x+1}$. 
We denote the four eigenstates of $H^y_{i_x}$ on rung $i_x$ as 
\begin{align}
	\ket{0}     & = \ket{\rm vac},                                                                & 
	\ket{1^{+}} & = \frac{ \ahatdag_{i_x,1} + \ahatdag_{i_x,2} }{\sqrt{2}}\ket{\rm vac}, \notag \\
	\ket{2}     & = \ahatdag_{i_x,2}\ahatdag_{i_x,1}\ket{\rm vac},                          & 
	\ket{1^{-}} & = \frac{ \ahatdag_{i_x,1} - \ahatdag_{i_x,2} }{\sqrt{2}}\ket{\rm vac},
\end{align}
where $\ket{\rm vac}$ denotes the vacuum on rung $i_x$.
The corresponding eigenenergies $E_y$ of $H^y_{i_x}$ are listed in \tref{tab:eigval}.
We then express  $\ahat_{i_x,i_y}$ and $ \ahatdag_{i_x,i_y}$ in terms of these eigenstates, plug them
into $H^x_{i_x, i_x+1}$ and obtain:
\begin{equation}
	\begin{split}
		- H^x_{i_x,i_x+1}/J_x  &=
		\ket{0; 1^{+}} \bra{1^{+}; 0} + \ket{0; 1^{-}}\bra{1^{-};0}   \\
		&  +
		\ket{2;  1^{+}} \bra{1^{+}; 2} + \ket{2; 1^{-}}\bra{1^{-};2} \\
		&  +
		\ket{1^{+}; 1^{+}} \bra{0; 2} - \ket{1^{-};1^{-}} \bra{0; 2}   \\
		&  +
		\ket{1^{+}; 1^{+}} \bra{2; 0} - \ket{1^{-};1^{-}} \bra{2; 0} \\
		&  + h.c.  \, .
	\end{split}
	\label{eq:Hxdiag}
\end{equation}
Here, $\ket{\alpha; \beta} \equiv \ket{\alpha}\otimes \ket{\beta}$ with $\alpha, \beta \in \set{0,1^{+},1^{-},2}$ denotes the
tensorproduct of the eigenstates on rungs $i_x$ and $i_x+1$.
The terms in the first two lines of \eref{eq:Hxdiag} correspond to just 
 an exchange of the eigenstates $\alpha \leftrightarrow \beta$ between the neighboring sites.
Thus we can identify the terms of the first line to drive the propagation of single bosons on top of the vacuum.
The second line can be seen as the propagation of a particle on top of a one-particle background, or alternatively,
a single hole in the background of filled rungs.

In contrast to the terms of the first two lines, the terms in the third and fourth row of
\eref{eq:Hxdiag} mix different eigenstates. 
If we imagine to start from a domain wall $\ket{\dots ; 2; 2; 0; 0; \dots}$, those  are the 
terms which ``create'' the single particle modes $\ket{1^{\pm}}$ at the border of the domain wall.
Subsequently, we would imagine these modes to propagate away to
the left as single-hole modes and to the right as single-boson modes.
Yet,  for the two-leg ladder all these mixing terms change the total energy $E_y$ from $0$ to
either $+ 2J_y$ or $- 2 J_y$. 
Thus, the creation is only possible via higher-order processes, which are suppressed with increasing $J_y/J_x$.
A term such as  $\ket{1^{+};1^{-}}\bra{2; 0}$ would not change the total energy $E_y$,
but such a term is not present in \eref{eq:Hxdiag} due to the conservation of total momentum $k_y$: it would
change from $k_y = 0+0 $ to $k_y = 0+\frac{\pi}{a}$.

To summarize, we argue that the $L_y=2$ ladder is special as it possesses the two extremal modes $k_y=0$ and
$\pi$ with a large energy $ E_y = \pm J_y$ for one particle on a rung. 
Note that energy conservation for large $J_y \gg J_x$ does not suppress the propagation of the modes
$ \ket{1^{\pm}} $ in the vacuum, but the creation of these modes at the edges of the initial blocks or a
domain wall.
As a consequence the current decays very rapidly as evidenced in \fref{fig:DeltaN}(a).

\begin{table}
	\begin{tabular}[t]{|l|l|r|l|} \hline
	\multicolumn{4}{|c|}{$L_y=2$ ladder} \\ \hline
	$N$ & $k_y\,[\frac{\pi}{a}]$ & $E_y\,[J_y]$ & state         \\ \hline \hline
	0   & 0         & 0       & $\ket{0}$     \\ \hline
	1   & 0         & -1      & $\ket{1^{+}}$ \\ \cline{2-4}
	    & 1         & 1       & $\ket{1^{-}}$ \\ \hline
	2   & 0         & 0       & $\ket{2}$     \\ \hline
	\end{tabular}
	\hfill
	\begin{tabular}[t]{|l|l|r|} \hline
	\multicolumn{3}{|c|}{ $L_y=4$ cylinder} \\ \hline
	$N$  & $k_y\,[\frac{\pi}{a}]$ & $E_y \,[J_y]$ \\ \hline \hline
	0; 4 & 0         & 0       \\ \hline
	1; 3 & 0         & -2      \\ \cline{2-3}
	     & 0.5       & 0       \\ \cline{2-3}
	     & -0.5      & 0       \\ \cline{2-3}
	     & 1         & 2       \\ \hline
	2    & 0         & -2.828  \\ \cline{2-3}
	     & 0.5       & 0       \\ \cline{2-3}
	     & -0.5      & 0       \\ \cline{2-3}
	     & 1         & 0       \\ \cline{2-3}
	     & 1         & 0       \\ \cline{2-3}
	     & 0         & 2.828  \\ \hline
	\end{tabular}
	\hfill
	\begin{tabular}[t]{|l|r|} \hline
	\multicolumn{2}{|c|}{$L_y=4$ ladder} \\ \hline
	$N$  & $E_y\,[J_y]$ \\ \hline \hline
	0; 4 & 0       \\ \hline
	1; 3 & -1.618  \\ \cline{2-2}
	     & -0.618  \\ \cline{2-2}
	     & 0.618   \\ \cline{2-2}
	     & 1.618   \\ \hline
	2    & -2.236  \\ \cline{2-2}
	     & -1      \\ \cline{2-2}
	     & 0       \\ \cline{2-2}
	     & 0       \\ \cline{2-2}
	     & 1       \\ \cline{2-2}
	     & 2.236  \\ \hline
	\end{tabular}
	\caption{Eigenenergies of a single rung. For a given particle number, degenerate levels are listed by their
	multiplicity.
}
	\label{tab:eigval}
\end{table}
\subsection{Larger cylinders and ladders}
We turn now to the cylinder and the ladder with $L_y = 4$.
The eigenenergies of $H^y_{i_x}$ on a single rung are listed in \tref{tab:eigval}. 
Giving an explicit expression for $H^x_{i_x,i_x+1}$ on an $L_y=4$ cylinder or ladder is not possible here, since it
contains too many terms. Nevertheless, we examine its structure. 
Similar to that for the two-leg ladder, we can distinguish between terms which just exchange the eigenstates
of neighboring rungs and terms which mix them.
As on the two-leg ladder, we associate the exchange terms with the propagation of modes.
Since $H^x$ contains only single-particle hopping, the exchange terms appear only 
between eigenstates with $N$ and $ N+1$ bosons on neighboring rungs.
Thus, to first order in $J_x/J_y$, a mode of $N$ bosons can propagate ``freely'' only in a background of
$N\pm 1$ bosons per rung.
By definition, all these exchange terms do not change the total energy $E_y$.

For the mixing terms, there is no restriction on the initial particle numbers on the neighboring
rungs. However, $H^x_{i_x,i_x+1}$ obviously preserves the total number of particle,
thus there are only mixing terms for 
$\ket{\dots N, N'\dots} \leftrightarrow \ket{\dots N\pm 1; N' \mp 1\dots}$.
The initial  melting of the edge thus happens via  a cascade 
of subsequent mixing processes. For example, consider 
\begin{multline}
	\ket{\dots 4; 4; 0; 0\dots} \rightarrow \ket{\dots 4; 3; 1; 0\dots}  \rightarrow
	\ket{\dots 4; 2; 2; 0\dots}  \\
	\rightarrow \ket{\dots 3; 3; 2; 0\dots} 
	\rightarrow \ket{\dots 3; 3; 1; 1 \dots} \, . 
	\label{eq:cascade}
\end{multline}
On the cylinder there are states with $E_y=0$ for any number of bosons per rung (see
\tref{tab:eigval}). 
This makes it plausible that cascades like (\ref{eq:cascade}) are possible without changing $E_y$ on the single rungs. 
Indeed, we find the corresponding terms in the expression for $H^x_{i_x,i_x+1}$ (not given here).
The initial edge of a block or domain wall can thus gradually melt into states with one particle per rung while
preserving the energy $E_y$.
This is confirmed by a strong peak in the momentum distribution function depicted in \fref{fig:cylNk}(f).
These additional $k_y = \pm \frac{\pi}{2a}$ modes with $E_y=0$, which are not present in the two-leg ladder, explain thus
the trend of a faster expansion seen as higher velocities in \fref{fig:cylladvcvr}.
Moreover, we stress that this process is independent of $J_y$, provided that other modes with $E_y
\neq 0$ are suppressed and our picture is applicable. Indeed, we find that the velocities in
\fref{fig:cylladvcvr} and currents (slopes) in \fref{fig:DeltaN}(b) are roughly independent of
$J_y/J_x$, even for moderate $J_y/J_x \gtrsim 0.6$.

On the other hand, on the four-leg ladder, there are no states with $E_y=0$ for one or three bosons on a
rung. It is thus immediately clear that there can be no mixing terms which preserve $E_y $ on
every rung separately.
Moreover, we find that there are also no mixing terms which create modes with opposite energy starting from
$E_y=0$ on both rungs.
As a consequence, the domain wall melting on the four-leg ladder requires higher-order processes, similar to the
two-leg ladder. However, the necessary intermediate energies $E_y= \pm 0.613 \times 2J_y$ are smaller than for the two-leg ladder,
such that these higher-order processes processes are more likely.
This is reflected in \fref{fig:cylladvcvr} by higher velocities for the four-leg ladder compared to the two-leg ladder.


\bibliography{references}

\end{document}